\documentclass[preprintnumbers,superscriptaddress,reprint]{revtex4-1}
\usepackage{textcomp}
\usepackage{amsmath}
\usepackage{amssymb}
\usepackage{graphicx}
\usepackage{color}
\usepackage{soul}
\newcommand{\ntu}
{\affiliation{School of Science and Technology, Nottingham Trent University, Clifton Lane, Nottingham NG11 8NS, UK}}

\newcommand{\mpi}
{\affiliation{Max Planck Institute for Dynamics and Self-Organization, Am Fassberg 17, 37077 G\"{o}ttingen, Germany}}

\begin{document}

\title{Load Dependence of Power Outage Statistics}

 \author{Soumyajyoti Biswas}
 \email{soumyajyoti.biswas@ds.mpg.de}
 \mpi
 
 \author{Lucas Goehring}
 \email{lucas.goehring@ntu.ac.uk}
 \ntu
 \date{\today}

 \begin{abstract}
 The size distributions of power outages are shown to depend on the stress, or the proximity of the 
load of an electrical grid  to complete breakdown. Using the data for the U.S. between 2002-2017, we show that the 
outage statistics are dependent on the usage levels during different hours of the 
day and months of the year. At higher load, not only are more failures likely, but the distribution of failure sizes shifts, to favor larger events.
At a finer spatial scale, different regions within the U.S. can be shown to respond differently
in terms of the outage statistics to variations in the usage (load). The response, in turn, corresponds to the
respective bias towards larger or smaller  failures in those regions. We provide a simple model,
using realistic grid topologies, which can nonetheless demonstrate biases  as a function of the
applied load, as in the data. Given sufficient data of small scale events, the method can
be used to identify vulnerable regions in power grids prior to major blackouts.

\end{abstract}
\maketitle

\section{Introduction}
Sustained and secure supply of power is a vital component of a prosperous society. Other essential services, such as water supply, medical infrastructure, communication, transport and so on are all dependent on the stability of their power supplies. Interruptions or outages in the power supply can have catastrophic consequences, as was witnessed in many
occasions (\textit{e.g.} August 14$^{th}$, 2003 in the U.S. and Canada; September 28$^{th}$, 2003 in Italy; and July 30$^{th}$, 2012 in India) \cite{list}.  Characterizations of power grid instabilities and outages, therefore, have been active topics of research for decades in engineering and physics communities (see \textit{e.g.}, \cite{anatomy,dobson2015,timme12,pahwa14,brummitt12,simonsen08,vaiman12,ji16}).  An operating power grid, particularly near its permissible level of capacity, can suffer from large outages triggered from small initial fluctuations or disturbances. For example, a software failure in an early warning management system \cite{report}, a falling tree on a line \cite{report_italy} or overloading by users \cite{blackout_india} caused the above mentioned
blackouts affecting, respectively, about 55, 56 and 620 million people. An amplified response to a small scale perturbation 
is a prominent signature of system-wide correlations developed near a critical point. 


A clear example of correlated response in power grids is the distribution function of the outage sizes, determined, for example, by the number of customers left unserved during an outage.  While a random failure probability would give an exponentially decaying distribution for outage sizes, in reality the probability $P(S)$ of an outage of size $S$ has a power law tail \cite{2007chaos}: $P(S)\sim S^{-\alpha}$, implying relatively higher probabilities for large outages. This is due to local correlations and causally connected cascades or avalanches of outage events.  Statistical analysis of those avalanches, particularly those demonstrating the universality of the exponent value $\alpha$ across different countries \cite{2007chaos}, has led to the identification of the dynamics of the power grid avalanches with that of self organized criticality (SOC) \cite{dobson00,dobson04}. Indeed, a connected set of objects (grid lines) having finite failure thresholds, with drive (customer demand) and dissipation (load unserved) is a suitable system for showing universal
collective behavior in a self-organized critical state.  Drawing such a parallel allows for investigation of power grid dynamics using the
standard tools of SOC developed over decades \cite{bak}. Furthermore, it also puts power grids in the generic class of driven dissipative systems 
having intermittent activities or avalanches with scale free size distributions. For example, this
is reminiscent of the Gutenberg-Richter-like law seen in various scales such as originally for the earthquake statistics \cite{gr}, also for stressed brittle solids \cite{main_kun}, 
sheared granular media \cite{takahiro} and so on.



While in the steady state the time series of the avalanches show scale free size distribution with universal exponent value, 
a common observation in these systems is the variation of this exponent value with the `load' on the system. Load is in
general considered here to be the relevant driving field, e.g. tectonic stress for earthquakes, compressive stress for 
fracture experiments and so on. 
Specifically, if only the events occurring at a higher stress are sampled, the magnitude of the exponent $\alpha$ is smaller than
what is obtained for events occurring at a lower stress. This was first observed for sheared rocks \cite{scholtz}
where the exponent decreases linearly with the differential stress. Subsequently, it was observed in other
failure dynamics \cite{amitrano}, including famously for the Gutenberg-Richter law exponent in earthquakes \cite{sch05,scholz15}.
Although the magnitudes of earthquakes follow universal scaling, in some regions the exponent value tends to 
be lower, signaling a higher risk of large earthquakes \cite{obara}.

Here we show that such lowering of size distribution exponent with increased load, i.e. customers' demands, also takes place for 
power outage statistics. Using the data for the outages in the U.S. between 2002-2017, we found that the size distribution
of the outages between night and day times, where the usage level approximately changes by 35\%, are significantly different. 
Such changes in the exponent are also observed for smaller regions in the U.S., where it can be
indicative of the relative risks of outages. Indeed there is a systematic variation of the exponent value with the load on
the grid for different hours of a day and different months of a year. We are able to reproduce this feature using a minimal 
model both for a realistic topology of the U.S. grid and other simpler topologies.
The load dependence of the size distribution exponent for outages opens a new path towards possible forecasting of 
large outage prone regions. For power grids, such identifications are very advantageous, as focused mitigation 
efforts (e.g. upgrading lines) can help in preventing large outages.
 
\section{Outages in the U.S. grid}

\begin{figure}[t]
\centering
\includegraphics[width=7.6cm]{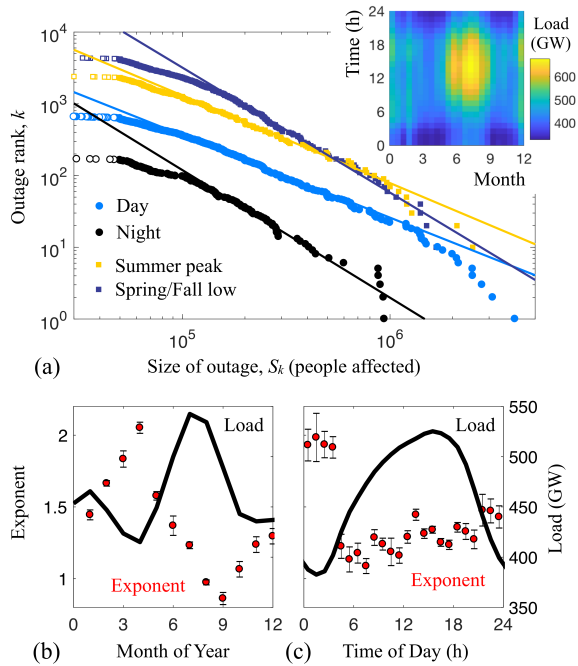}
\caption{The sizes of large-scale outages in the U.S. follow a power-law distribution, whose exponent changes with the load on the grid at the time of outage. (a) We demonstrate this by dividing outages into those occurring during the day and night, or the summer and off-peak/winter periods.  The lower load cases (night and winter) show power-law distributions with steeper distributions than the higher load cases (day and summer).  The summer/winter plots are shifted 10x up the y-axis, to aid visibility. The inset shows the national electricity consumption at different times (Pacific time) and months, for 2016.  This robust anti-correlation between the load and exponent can be seen if the data are further subdivided according to (b) month (3-month rolling average) or (c) time of day (3-hour rolling average).}
\label{fig1}
\end{figure}

The size distribution of the power outages in the U.S. has been studied, both in terms of the power left unserved and the number of customers affected \cite{clauset2009,dobson2015,hines09}. In fact, these quantities vary almost co-linearly, except in a few instances involving load shed affecting \textit{e.g.} one customer, such as may be the case for a large industrial facility. The two metrics also show power law size distributions and are more or less equivalent in terms of the exponent values \cite{hines09}. For example, the cumulative distribution function for the number of customers affected during blackouts has been reported to follow a power law, with exponents variously estimated in the range of 0.8 - 1.3 \cite{clauset2009,hines09,dobson2015}.  Similar studies for outages in Sweden \cite{hol06}, Norway \cite{bakke06}, New Zealand \cite{ancell05} and China \cite{weng06} also show scale-free size distributions of power outages.  Here we show that the exponent value of such distributions depends on the load carried by the power grid, at the time of failure.

For events with a power-law size distribution, the probability of an event of size $S$ scales as $p(S) \sim S^{-\alpha}$.  When these events are arranged in descending order of magnitude, the resulting rank-plot will follow $k\sim S_k^{-B}$, where $S_k$ is the $k$-th largest event, and the exponent $B = \alpha-1$ (see Methods).  

In Fig. \ref{fig1}(a) the rank plots, or cumulative size distributions, are given for the subsets of power outages occurring respectively during the day (08:00-20:00, local times), night (22:00-04:00), summer (July-October), or winter (October-May).   These periods were chosen to correspond with the times of peak and off-peak loads, as measured in the national electricity demand during 2016, and shown in the inset to \ref{fig1}(a).   All data are taken from the public reports of the U.S. Energy Information Administration \cite{USEIA}, which lists outage events affecting more than 50,000 consumers, or resulting in a load shedding of more than 300 MW, as well as the hourly electricity demand.   The outage data used covers 1193 events in the years 2002-2016.

A power law fit of the whole data set gives an exponent $B = 1.30\pm0.02$, consistent with previous reports \cite{clauset2009,dobson2015,hines09}.  However, with a day/night division these outages split into a shallower daytime distribution with $B = 1.15\pm 0.03$ and a steeper nighttime distribution with $B = 1.78 \pm 0.02$.  While it is known that there are fewer outages at night, than in the day \cite{hines09}, this result shows that those outages that do occur at night are generally also much less severe.  Similarly, if the data are split seasonally, we find an exponent of $B = 1.22\pm 0.04$ in the months of peak summer usage, but an exponent of $B = 1.74\pm0.04$ during the off-peak winter months.   

To show that there is a significant relationship between the load on the grid and the exponent value of the outage size 
distribution, we have considered outages in rolling 3-hour time windows.  Fig. \ref{fig1}(b) shows the variations of the exponent $B$ and the load for different hours of the day. Similarly, Fig. \ref{fig1}(c) shows the load and exponents for different months of the year, using a three month window. 
For both cases an anti-correlation between the load and the fitted exponent value can be seen.    Like a variety of other driven disordered systems, including earthquakes \cite{sch05,scholz15}, laboratory scale fracture \cite{amitrano}, and sheared granular systems \cite{riviere2017}, we find that a higher load is associated with a smaller $B$ value, and hence a more extreme distribution of events.

\begin{figure*}[tb]
\centering
\includegraphics[width=14.6cm]{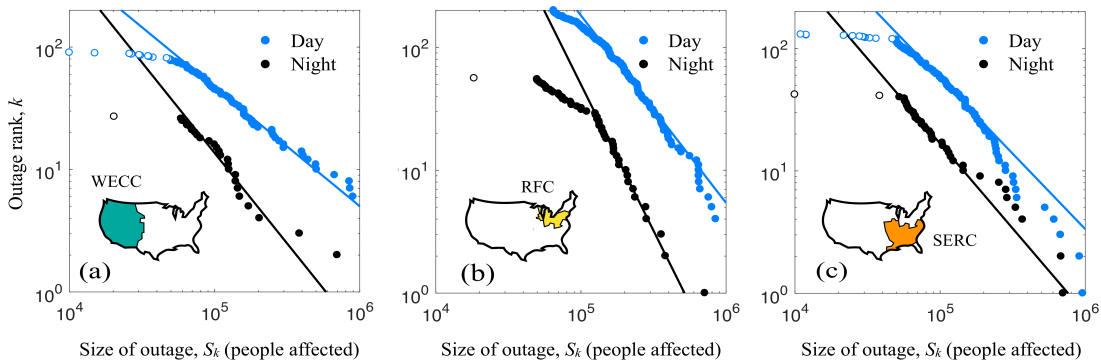}
\caption{Day-night variations in outage size distributions are also seen regionally.  Outages in the U.S. are divided according to their governing Regional Reliability Council (RRC).  Shown are the rank size distributions for the three regions with the most reported events, (a) WECC, (b) RFC and (c) SERC, divided between day (10:00 - 20:00, local time) and night (00:00-06:00).   In each case the night-time outage distribution is steeper than the day-time distribution, indicating that outage events are generally more severe at higher-load times.}
\label{fig2}
\end{figure*}

%

So far we have shown the temporal variation of the outage size distribution exponent over the entire data. However, 
to identify the vulnerable areas or dangerous hot-spots on a grid, it is important to analyze such load dependence on different spatial
segments as well. The U.S. grid is divided between 10 electricity regulatory authorities \cite{map}. We chose the 
largest three in terms of the number of events and performed the similar analysis of splitting the data between day and night
and
find a similar variation in the exponent values (see Fig. \ref{fig2}).  
This establishes that our method can be useful for identification of a vulnerable sub-volume of a larger grid.  
For that, however, one needs to calibrate the variation in exponent with risk of large outage, which requires a large volume
of fine grained data. With the present data, requiring more than 50,000 people or loss of 300MW of load to get reported, such analysis 
is not feasible. 

\section{Model}

We now explore the load-dependence of power grids \textit{via} a simple network model, with different topologies and loading conditions.  There are several approaches to modeling the dynamics of a power grid, including examples of networks obeying circuit laws \cite{chen05,rios02,kirschen04,simonsen08}, sometimes incorporating phase information \cite{timme12,yang17}, as well as more abstract models \cite{demarco, stubna03}, alongside a large volume of literature on failures in complex networks in general (see e.g. \cite{boccaletti03,buldyrev10,goh01,panzieri}). 
Here we use a model of the power grid similar to that studied in Refs. \cite{pahwa10,yagan1,yagan2}, and demonstrate how the observed U.S. outage data match generic features of the load-dependence of outage statistics.  

Specifically, we consider a set of elements, or nodes, having finite failure thresholds. The elements are either arranged on a regular grid, or connected to each other by the topology given in Ref. \cite{watts98}, which simulates the Western Interconnection of the U.S. grid. The thresholds $\sigma_{th}^i$ and loads $\sigma_l^i$ of the $i$-th element are related by
\begin{equation}
\sigma_{th}^i=\sigma_l^i+s\epsilon_i.
\end{equation}
We assign a random load $\sigma_l^i$ to each element, from a uniform distribution between zero and one.  The second term on the right hand side provides a buffer or redundancy for the elements.
This ensures that the capacity of an element is always higher than its initial load.   The random variables $\epsilon_i$ are also chosen from a uniform distribution on $[0,1]$. Therefore, on average, the network carries a load of $1/(1+s)$, relative to its maximum capacity.

The dynamics of the model follows from randomly choosing an element and equating its threshold to its load thereby triggering a failure event. This can happen due to external
causes (storm, vandalism etc.) on a grid or due to sudden surge in demand among the customers and so on. The load carried by that element
now has to be redistributed among the remaining surviving elements. That may, in turn, cause some of those elements to break,
triggering an avalanche. We can quite reasonably assume a separation of time scales between successive triggering events and 
internal redistribution of loads. Therefore, during an avalanche the total load remains constant and the only dynamics is the
redistribution of loads in successive steps following a breakdown. An avalanche is then the number of elements breaking until a stable configuration is reached.
 After an avalanche, all the elements are again restored with another random threshold and
load chosen from the same distributions. This allows an average over disorder in the system.
Clearly, the value of $s$ will determine the relative stress on an element with respect to its carrying capacity. Higher the $s$, lower is the relative stress.

It is important to note, however, that due to the long-range nature of the correlation developed in the system, a failure somewhere in the grid can trigger an avalanche or cascade failure at a different location. Specific examples of such events include 
the August 10, 1996 outage \cite{96outage}. Therefore, one version of our model is long range, taking into account the long range nature of electrical current variation, so that 
a local disturbance can indeed trigger a remote event. In another version, we take the exact topology \cite{watts98} of the Western Interconnection.



\begin{figure}[tb]
\centering
\includegraphics[width=7.6cm]{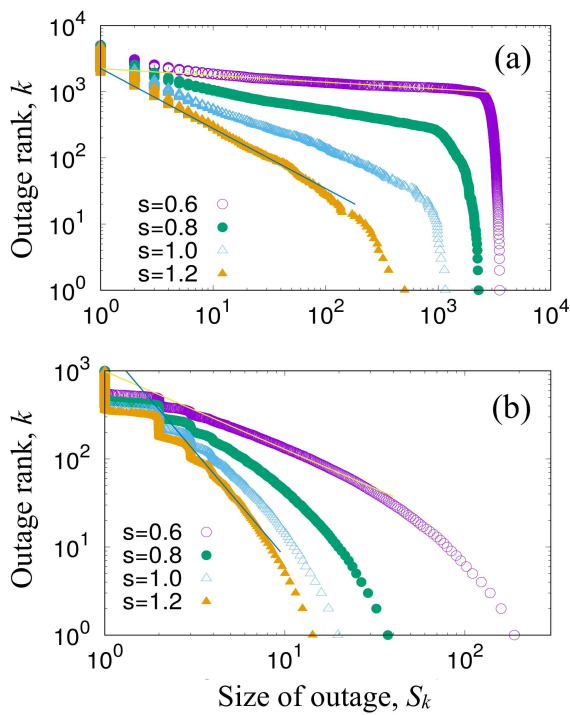}
\caption{The rank-plot of the avalanches observed for the two-dimensional version of the model with
power-law load redistribution for various values of stress $s$. In the range shown, the exponent value
changes from $0.62$ to $1.80$, which include the range seen in the data (Fig. \ref{fig1}).}
\label{ava_stress_rank}
\end{figure}
While the statistics of the avalanches can depend on the particular topology considered, the qualitative observation of lowering
of the size distribution exponent (or that of the rank plot) should remain valid independent of it. To begin with,
we consider the two-dimensional square lattice network. Following a local failure, the load carried by the element is redistributed 
to the entire remaining network. However, the load sharing depends with distance from the failure point in a power law
$1/|x-x^{\prime}|^{\gamma}$ (considering periodic boundary conditions). For the simulations here we have taken $\gamma=2$ in keeping with the similar dependence of current 
flow in the random fuse model \cite{lucilla}. The resulting avalanches are recorded over time and the above mentioned rank-plots are shown in Fig. \ref{ava_stress_rank}
for different values of the relative stress $s$. The exponent varies
significantly within the range of $s$ studied. Particularly, for the range $s=0.55-0.9$, the exponent value changed from $0.88-1.87$. This covers the 
observed range ($1.2-1.80$) in the data for the whole country or that of its parts ($1.1-1.4$).

\begin{figure}[tb]
\centering
\includegraphics[width=7.6cm]{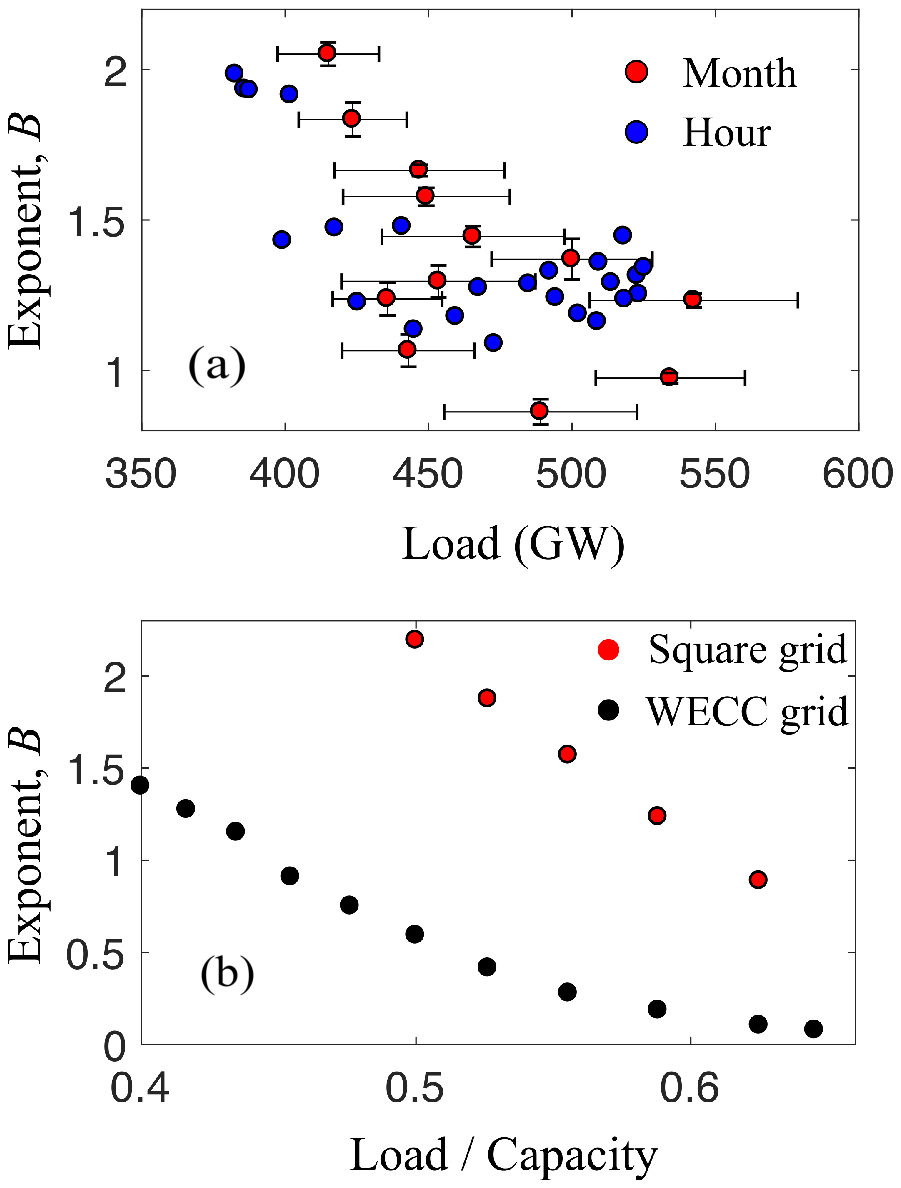}
\caption{The top figure shows the scatter plot of the exponent values from Fig. \ref{fig1} with the corresponding average load showing representative error bars. A linear trend of decay in the exponent value is observed.
The figure at the bottom is the rank-plot of the avalanches observed in  the model with
same topological structure as the North-Eastern power grid \cite{watts98}. For various values of stress $s$ the exponent value
changes from $0.1$ to $1.5$, which has substantial overlap with the range seen in the data (Fig. \ref{fig1}).}
\label{ava_topology}
\end{figure}
The topology of the power grid is also an important factor in its dynamics. The characterization of the network properties
is a debated issue \cite{pagani13}, with claims of small-world and scale-free properties in the topologies of power grids in various countries. 
Nevertheless, the qualitative feature of stress-dependence as discussed above, are not expected to change with the topology of
the grid.  To have a more realistic feature in our model, we have studied the model in the exact topology of the North-Eastern grid
as reported in Ref. \cite{watts98}. The load sharing in this case is confined only among the connected neighbors. The resulting 
statistics (Fig. \ref{ava_topology}) of the outages for various values of stress still has a substantial overlap in the observed exponent values.

\section{Discussion and Conclusions}
The intermittent dynamics of the power grid outages and its association with self-organized criticality 
is known for over a decade now. Characteristic signatures, including scale-free size distribution of 
the outage sizes are seen for outages in different countries, for example, the U.S., Sweden, Norway, China. 
This connection of power grids with self-organized criticality enables comparisons of statistics of 
power outages with other similar systems, for example, that of earthquakes. 

Here we show that like in earthquakes and several other driven disordered systems, for power grids as well 
the exponent value of 
the size distribution of the avalanches, which in this case is the number of customers affected in an 
outage, depends on the load on the system. Particularly, for higher values of the load, the magnitude of
the exponent becomes smaller, indicating a relatively increased probability for outages of higher sizes. 
We demonstrated this anti-correlation between the load on the grid and the exponent value of the outage size
distribution by analyzing the data for different months of the year and different hours of the day (Fig. \ref{fig1}).
This property is also valid for sub-regions, for example in several Reliability Commission areas in the U.S.  
The scatter plot of the exponent values with the average load on the grid at the time of outage show
a clear signature of the decay of the exponent value with the load (Fig. \ref{ava_topology}), which is almost linear, as
in the case of earthquakes \cite{scholtz}.   

Given the generic nature of the anti-correlation between load and avalanche size distribution exponent, 
we attempted to use a toy model for power grids to explore the effect of load dependence on the outage 
size distribution. The model is defined as a collection of nodes, each carrying a load lower than their randomly
assigned capacity,  connected in certain topologies. 
We tested the results in a square lattice topology with distance dependent load redistribution, as well as 
the exact topology of the Western Interconnects (Fig. \ref{ava_topology}).   Depending on the average load on the system, the model
reproduces the load dependence of the avalanche size distribution as seen in the model for various topologies.

In conclusion,  we have demonstrated that scale free variation of the outage size distribution in power grids depends on the
load on the grid at the time of outage. The variation of the exponent, an almost linear decay of the exponent 
value with the load, is similar to that found in earthquake size distributions. Given sufficient resolution
of the outage data, the method can be used to identify vulnerable regions of power grids, as is done for
statistical predictions of earthquakes.

\section*{Methods}
Data are collected from the U. S. Energy Information Administration website \cite{USEIA}, over the period 2002-2016, inclusive.  There are 1193 reported forced (i.e. unplanned) outages in this period affecting known, non-zero numbers of customers, which we consider.  Outage times are given according to the appropriate local time zone.  The load values used are for 2016, and reported nationally using the Pacific time (PT) zone as a reference \cite{USEIA}.  Hourly load data was taken from the 1st and 15th of every month, avoiding weekends and holidays (specifically, using January 4th/15th; May 2nd/13th; and October 3rd/14th), when load patterns would be different.  Daily average loads were collected on each day throughout the year.   Averages and standard deviations of the load were calculated from these data for each window of hours or months used.  

For fitting the outage size distributions, we use rank plots.  The events $S$ can be arranged in the descending order of their sizes:
\begin{equation}
 S_1 \ge S_2 \ge S_3 \dots \ge S_n. \nonumber
\end{equation}
The $k$th ranked element has size $S_k$.  For events with a probability distribution $p(S)$, the number of events having size greater than or equal to $S_k$ is
\begin{equation}
  \int_{S_k}^{\infty} p(S)dS=k.
\end{equation}
If $p(S)\sim S^{-\alpha}$, one then has
\begin{equation}
  k\sim S_k^{(1-\alpha)} = S_k^{-B}.
\end{equation}

The ranked data was fit in two ways.  First, a maximum likelihood estimator (MLE) method \cite{clauset2009} was applied to measure the exponent $B$, and an event size cutoff.  Second, we applied a least-squares fit to data that were binned to have equal widths on a log-scale.  For this an \textit{a priori} cutoff is required.  This is taken as 50,000 (the requirement for reporting) or somewhat above, when there are other signatures of under-reporting (\textit{e.g.} kink at 100,000 consumers in Fig. \ref{fig2}(b)).  Error estimates were checked by repeating fits on randomly subsampled data sets (100 trials on half-sampled data); the resulting spread in exponents is consistent with the stated fit errors.  While the precise exponent values differ depending on the method of fitting used, they show same trends with load.  
Fits given in the manuscript are least-squares fits.

\end{document}